\documentstyle[12pt]{article}

  \def\cite#1{[\ref{#1}]}

  \def\citm#1#2{[\ref{#1}--\ref{#2}]}

\def\ben{\begin{enumerate}}  \def\een{\end{enumerate}}
\def\beq{\begin{equation}}   \def\eeq{\end{equation}}
\def\bea{\begin{eqnarray}}  \def\eea{\end{eqnarray}}
\def\nn{\nonumber}
\def\noi{\noindent}

\def\lsim{\raise0.3ex\hbox{$<$\kern-0.75em\raise-1.1ex\hbox{$\sim$}}}
\def\gsim{\raise0.3ex\hbox{$>$\kern-0.75em\raise-1.1ex\hbox{$\sim$}}}

\begin{document}

\vbox to 1 truecm {}
\begin{center}
{\large \bf Decoherence : An Irreversible Process} \par

\vspace{1 truecm}

{Roland Omn\`es}\footnote{email : roland.omnes@th.u-psud.fr}\\
{\it Laboratoire de Physique Th\'eorique}\footnote{Unit\'e Mixte de Recherche
UMR 8627 - CNRS }\\    {\it Universit\'e de Paris XI, B\^atiment 210, 91405
Orsay Cedex, France}
\end{center}

\vspace{2 truecm}
\begin{abstract}
A wide-ranging theory of decoherence is derived from the quantum theory of
irreversible processes, with specific results having for their main
limitation the assumption of an exact pointer basis.
   \end{abstract}

\vspace{2 truecm}

\noi LPT Orsay 00-74 \par
\noi Mai 2001
 
\newpage
\pagestyle{plain}
Decoherence has become widely recognized as an essential step for
understanding and interpreting quantum mechanics. It has been mainly
investigated on solvable models, which give most of what we know about the
effect~; but though some of these models are considered as realistic
in a definite situation (for instance in quantum optics
\cite{1r}), they remain in most cases rather far from reality. Now
that the importance of decoherence is acknowledged, it certainly requires
a wider theory, and forthcoming investigations
will also need more precise results (particularly in the research on
quantum computers, where decoherence is expected to be the main
obstacle and must be controlled quantitatively). \par

A simple starting point for such a theory consists in noting that
decoherence is an irreversible process, so that one can apply the rather
general theories for such processes. It is generally agreed
that the best theories of that sort rely on the
so-called ``projection method'' \citm{2r}{5r}. There have already been some
attempts to apply it to decoherence \cite{6r}, but the results
in the present Letter improve on them on at least three significant
points~: \par

\ben
\item One can always define an average part in the
coupling of the system with the environment, and when that part is removed
the remaining coupling consists only of fluctuations~; one can therefore
extend considerably the range of perturbation theory. \par

\item A rather general master equation for
decoherence is obtained (Eq. (\ref{12e}) below). \par

\item Precise quantitative equations are obtained for
decoherence, at least when an exact ``pointer basis'' exists \cite{7r}.
\een

\centerline{\hbox to 3 truecm{\hrulefill}}

\vskip 3 truemm

As far as the overall framework is concerned, one assumes as usual that
the system under consideration can be split into a ``collective''
subsystem with hamiltonian $H_c$ and an environment having a very large
number of degrees of freedom with hamiltonian
$H_e$. The two systems are coupled, and the full hamiltonian is $H 
=H_c \otimes I_e +
I_c \otimes H_e + H_1$, the coupling $H_1$
allowing energy exchanges and other mutual influences (including
decoherence) between the collective system and the environment. The full
density operator $\rho$ evolves according to the basic equation

\beq
\label{1e}
\dot{\rho} = -i [H, \rho] \quad .\eeq

\centerline{\hbox to 3 truecm{\hrulefill}}

\vskip 3 truemm

The main features of the projection method for a typical irreversible
process are as follows. A (countable or not) set of
independent ``relevant observables'' $A^i$ (including the identity $I$) is
selected. Their ``exact'' average values, resulting from the exact density
operator $\rho (t)$, are denoted by $a^i(t)$. A time-dependent test density
operator $\rho_0(t)$ is then introduced with the assumptions (i) that 
it gives the
exact average values $\{a^i (t)\}$ for the relevant operators $\{A^i\}$
(so that one might know them if $\rho_0$ is known), and (ii) that its
information content is minimal. It must be of the form

\beq
\label{2e}
\rho_0 = \exp \left ( - \lambda_i \ A^i \right ) \quad ,
\eeq

\noi with Lagrange parameters $\lambda_i$~; summation over repeated indices
is assumed as usual, and $\rho_0$ is normalized since the identity $I$
belongs to the set of relevant observables. Auxiliary ``densities'' (or
more properly trace-class operators) are then defined by $s_i = \partial
\rho_0/\partial a^i$. They satisfy the orthogonality properties,

\beq
Tr \left ( s_i \ A^j \right ) = \delta_i^j \quad ,
\label{3e}
\eeq

\noi which amount essentially to $\partial a^j/ \partial a^i = \delta _i^j$.
\par

The theory makes use of ``superoperators'', acting linearly on a
trace-class operator to yield a similar operator. For instance, Eq.
(\ref{1e}) can be written conventionally as $\dot{\rho} = {\cal L} \rho$,
where ${\cal L}$ is the so-called Liouville superoperator. Another
superoperator is defined in the projection approach by

\beq \label{4e}
{\cal P} = s_i \otimes A^i \quad ,
\eeq

\noindent which means that  ${\cal P}$ acts on a trace-class operator $\mu$
to give ${\cal P}{\mu} = Tr (A^i \mu ) s_i$. Eq. (\ref{3e}) implies the
projection property ${\cal P}^2 = {\cal P}$. A relevant density 
operator is defined as $\rho_1
= {\cal P}{\rho}$. It also yields the exact quantities $\{a^i\}$ as 
average values of the relevant
operators $\{A^i\}$, in view of Eq. (\ref{3e}). Denoting by ${\cal 
J}$ the identity superoperator, one also
introduces the superoperator ${\cal Q} = {\cal J} - {\cal P}$ which 
satisfies the same projection property ${\cal
Q}^2 = {\cal Q}$. Denoting ${\cal Q}{\rho}$ by $\rho_2$ (so that 
$\rho = \rho_1 +
\rho_2)$, and applying ${\cal P}$ and ${\cal Q}$ to both sides of Eq.
(\ref{1e}) one obtains evolution equations for $\rho_1$
and $\rho_2$~:

\beq
\label{5e}
\dot{\rho}_1 = {\cal P} {\cal L} {\cal P} \rho_1 + \dot{\cal P} {\cal P}
\rho_1 + {\cal P} {\cal L} {\cal Q} \rho_2 + \dot{\cal P} {\cal Q} \rho_2
\quad , \eeq

\beq
\label{6e}
\dot{\rho}_2 = {\cal Q} {\cal L} {\cal Q} \rho_2 - \dot{\cal P} {\cal Q}
\rho_2 + {\cal Q} {\cal L} {\cal P}\rho_1 - \dot{\cal P} {\cal P} \rho_1
\quad .\eeq

\centerline{\hbox to 3 truecm{\hrulefill}}

\vskip 3 truemm

When applying the projection method to decoherence, it will be convenient
to introduce a commuting set of collective observables $X$ whose
eigenvalues $x$ are either discrete or continuous. The ``relevant
observables'' $\{ A^i\}$ are chosen to consist of the identity $I$, 
the environment
hamiltonian $I_c \otimes H_e$ and the collective observables $(|x> 
\pm |x'>)\cdot (<x|
\pm <x'|) \otimes I_e$ and $(|x> \pm i |x'>) \cdot (<x| \mp i <x') 
\otimes I_e$ for every pair $(x,
x')$ of eigenvalues of $X$. One can use more simply the set of non-hermitian
operators $A^{xx'} = |x> <x'|\otimes I_e$, which clearly provide a basis
for the collective observable. The test density operator
(\ref{2e}) takes then the simple form

\beq
\label{7e}
\rho_0  = \rho_c
\otimes \rho_e \quad , \eeq

\noi where $\rho_c$ turns out to be the familiar reduced density operator
for the collective subsystem and $\rho_e$ is, at least formally, a
normalized density operator for the environment as if it were in thermal
equilibrium~:

\beq
\label{8e}
\rho_c = tr\ \rho \quad , \qquad \rho_e = \exp \left (- \alpha - \beta H_e
\right ) \quad . \eeq

\noi Throughout I denote a partial trace over the environment
by $tr$ and a full trace by $Tr$. The time-depending parameters $\alpha$
and $\beta$ are chosen so that $\rho_e$ is normalized and the ``exact''
average value $E$ for the environment energy is obtained from it.
It should be stressed that this expression of $\rho_e$ does not mean that
the environment is in thermal equilibrium~; it means only that one
does not need to know more than the average environment energy in order to
obtain collective quantities, including the reduced density
operator. \par

The first part of the calculation consists in obtaining algebraically 
obtaining the
auxiliary densities $s_i$, which are given here for convenience (using the
notation $s(A^i)$ in place of $s_i$)

\begin{eqnarray*}
&&s\left ( |x> <x'| \otimes I_e \right ) = |x'> <x'| \otimes \rho_e \quad
,\\
&&s\left ( I_c \otimes H_e \right ) = I_c \otimes \left ( \rho_e 
\left ( H_e - E \right ) \right ) \Delta^{-2}
\quad ; \quad s(I) = - E \ I_c \otimes \left ( \rho_e \left ( H_e - E 
\right ) \right ) \Delta^{-2}
\end{eqnarray*}

\noi with $\Delta^2 = tr (H_e^2 \rho_e) - E^2$, so that acting
on a trace-class operator $\mu$, the projection superoperator ${\cal P}$
gives

\beq
\label{9e}
{\cal P}{\mu} = (tr \ \mu ) \otimes \rho_e + I_c \otimes \left ( 
\rho_e \left ( H_e - E \right ) \right )
\Delta^{-2} \left \{ Tr \left ( H_e \ \mu \right ) - E\  Tr \ \mu \right \}
\quad .  \eeq
 
\noindent (In particular, $\rho_1 = \rho_0$). \par

Then comes an important trick. It is very convenient to introduce an average
collective coupling

$$\Delta H_c = tr \left ( H_1 \cdot I_c \otimes  \rho_e \right ) \quad ,$$

\noi which is a collective operator representing a collective
effect of the environment (for instance the action of pressure in the case
of a gaseous environment). It is generally important, though equal to 
zero in a few special cases
(matter-radiation coupling, nuclear magnetic resonance, and some 
oscillator models). The remaining part of the
coupling  $H'_1 = H_1 - \Delta H_c \otimes I_e$, consists presumably 
in most cases of small fluctuations, which
can be considered as perturbations of the hamiltonian $H_0 = (H_c + 
\Delta H_c) \otimes I_e + I_c \otimes H_e$. To
second order in $H'_1$, Eqs. (\ref{5e}-\ref{6e}) become explicitly

\beq
\label{10e}
\dot{\rho}_c = - i [H_c, \rho_c ] - i \ tr [H_1, \rho_2] \quad ,
\eeq

\beq
\label{11e}
\dot{\rho}_2 = - i [H_0, \rho_2] - i [H_1, \rho_0 ] + i \ tr
[H_0, \rho_2] \otimes \rho_e \quad . \eeq

\noi One can solve Eq. (\ref{11e}) for $\rho_2$ as a function of $\rho_0$,
using first-order perturbation theory and assuming for convenience that the
environmement is in thermal equilibrium at an initial time $t = 0$ so that
$\rho_2(0) = 0$. Inserting the resulting expression for $\rho_2$ into Eq.
(\ref{10e}), one obtains the rather simple and fundamental ``master
equation''

\beq
\label{12e}
\dot{\rho}_c = - i [H_c + \Delta H_c , \rho_c] - \int_0^t dt' \ tr \left \{
\left [ H'_1, U [H'_1, \rho_0 (t') ] U^{-1} \right ] \right \} \quad ,
\eeq

\noindent where $U = \exp (-i H_0(t - t'))$. \\

\centerline{\hbox to 3 truecm{\hrulefill}}

\vskip 3 truemm

Though already known when the full coupling $H_1$ is weak, Eq. 
(\ref{12e}) has a much wider range of valid since
it holds for a fluctuating $H'_1$ $^{\rm (a)}$. The last step in the 
calculation
consists in writing down explicitly the master equation, explicit
expressions being obtained when the basis $|x>$ is an exact pointer
basis or, more precisely, when $H'_1$ and $X$ commute so that $H'_1$ is
diagonal in the $|x>$ basis and behaves like an operator $V(X)$ in the
environment Hilbert space (more explicitly, introducing eigenvectors
$|k>$ of $H_e$ with eigenvalues $E_k$, one has $<x,k|H'_1|x',n> = \delta (x
- x') V_{kn}(x))$. The existence of a pointer basis will be assumed
from here on. \par

One can then define a microscopic distance (abbreviated by $\mu D)$
between two points $x$ and $x'$ as a distance $|x - x'|$ where the
(quantum) first term in the right-hand side of Eq. (\ref{12e}) dominates
the value of $\dot{\rho}_c(x, x')$. A small macroscopic distance ($SMD$) will
be one for which the second (decoherence) term in the right-hand side
dominates, although $x- x'$ is still macroscopically small. I will assume
furthermore that $V(x) - V(x')$ depends linearly locally on $x - x'$ with a
``slope'' $V'$ over small distances ($\mu D$ and $SMD$), which are the
only distances of interest in applications. The second (decoherence)
term in the right-hand side of the master equation (\ref{12e}) becomes
then, for $\mu D$'s and $SMD$'s,

\beq
\label{13e}
-\int_{0}^t dt' \int dx_1 \ dx'_1 \ K(x, x';x_1,x'_1;t-t') \rho_c(x_1,
x'_1, t') \eeq

\noi where the kernel is

  \bea
\label{14e}
K &=& \sum_{n,k,N,M} (x-x') \left [ (x_1 - x'_1) \cosh \beta
\omega_{kn} /2 + (x_1 + x'_1 - x - x') \sinh \beta \omega_{kn} / 2 
\right ] \nn \\
&&\times <x|N > <N |x_1> <x'_1|M> <M |x'> \exp (i \Omega_{MN}
{\tau}) \nn \\
&&\times |V'_{kn}|^{\, 2} \exp (-i \omega_{kn} {\tau}) 
\bar{\rho}_{nk} \quad . \eea

\noi By $|N>$, $|M>$, one denotes eigenvectors of $H_c + \Delta H_c$ with
eigenvalues $E_N$, $E_M$, and $\Omega_{MN} = E_M - E_N$, similarly 
$\omega_{kn} =
E_k - E_n$ and $\bar{\rho}_{nk} =  \exp [- \alpha - \beta (E_k +
E_n)/2)$, and ${\tau} = t - t'$. \par

This expression has many interesting consequences~:

\ben
\item The results of previous models can be recovered, often resulting in
a simpler kernel~; for instance, models with an environment consisting of a
collection of two-states systems \cite{8r} (one system for each $\omega$), or
harmonic oscillators \cite{9r}. Decoherence by collisions with
an environment of molecules or photons \cite{10r} is best obtained by using
plane-wave states for $n$ and outgoing scattering states for $k$.

\item Both terms involving $\cosh \beta \omega /2$ and $\sinh \beta \omega
/2$ in Eq. (\ref{14e}) are significant at low temperature and $\mu D$'s. This
case will be presumably important for future technology when decoherence
and quantum coherence compete.

\item The interpretation of quantum measurements, with suppression of
macroscopic superpositions (as in the Schr\"odinger cat problem), is
mainly concerned with macroscopic values of $(x - x')$, or $SMD$'s.
One can then put $x_1 = x$ and $x'_1 = x_1$, as can be shown easily when
$H_c = P^2/2m$ by means of Fourier transforms.
The case $H_c = P^2/2m + W(x)$ requires a more elaborate justification
using  coherent collective states or microlocal analysis, but one always
obtains for $SMD$'s

\bea
\label{15e}
K &\simeq& \sum_{n,k} \delta (x - x_1) \delta (x'- x'_1) (x - x')^2 
\cosh (\beta
\omega_{kn} /2 ) \nn \\
&&\times |V'_{kn}|^2 \exp \left [ - i \omega_{kn} (t - t') \right ] 
\bar{\rho}_{nk} \eea

\item At high enough temperature and when retardation in
$t-t'$ is neglected, the decoherence term at $SMD$'s becomes simply
$- \mu (x - x')^2$, with a decoherence coefficient

\beq
\label{16e}
\mu \simeq - \sum_{n,k} |V{'}_{kn}^{\, 2}| i (\omega_{kn} - i0)^{-1} 
\bar{\rho}_{nk} \simeq \pi \sum_{n,k}
|V'_{kn}|^2 \bar{\rho}_{nk} \delta (\omega_{kn} ) \eeq

\item If $P$ is the momentum canonically conjugate to $X$, one finds
easily that the term in $\cosh \beta \omega /2$ in Eq. (\ref{14e}) does not
contribute to $<dP/dt>$, i.e. to damping. The term in $\sinh \beta \omega
/2$ gives on the other hand gives a damping

\beq
\label{17e}
< dP/dt > \ = {\rm forces} \ - \int_0^t D(t-t') <P(t')> dt'
\quad ,
  \eeq

\noi with \hskip 2 truecm $D({\tau}) = 2 m^{-1} \sum\limits_{n,k} 
|V'_{kn}|^2 \bar{\rho}_{nk} e^{-i
\omega {\tau}} \sinh (\beta \omega_{kn} /2) \omega^{-1}_{kn} \quad .$

This simple result results from several steps~: The action of
$P = - i \partial / \partial x$ on the kernel $K$ (in Eq. (\ref{14e}))
removes the factor $(x - x')$ since other terms in the derivative
containing this factor have zero average. The
integral over $x$ of $<x|N > < M |x>$ gives simply $\delta_{NM}$. A term
such as $x <x|N><M|x> \exp (i \Omega {\tau})$ is expressed as
$<M |U_c^+ X U_c|N >$, with $U_c = \exp (- i H_c {\tau})$. One can then use
the equation

\beq
\label{18e}
[X, U] = UP{\tau}/m \quad ,
\eeq

which is exact for $H_c = P^2/2m$ and valid up to higher orders in $\hbar$
in the presence of a collective potential. The factor ${\tau}$ in Eq.
(\ref{18e}) is finally removed by integrating by parts on ${\tau}$.
Other more direct methods also exist for evaluating damping, and
they agree with the result. \par

\qquad When retardation effects are again neglected, the damping term in
Eq. (\ref{17e}) becomes $-\gamma <P>$ and, at high enough temperature, one
recovers the well-known relation $\mu = mkT\hbar^{-2}$, where Planck's
constant has been reintroduced. One might also derive the familiar
expression $-\gamma /2 (x-x')(\partial / \partial x - \partial / \partial
x')\rho_c(x, x')$ for damping with similar approximations.
\een

To conclude, the main limitation of the final results following Eq.
(\ref{12e}), is the assumption of an exact pointer basis. Such bases are
known to exist for a strictly mechanical collective system (the $X$'s being
the position coordinates of coarse-grained pieces of matter in the
macroscopic system), or in the case of SQUID loops \cite{7r}. More
generally, approximate diagonalization occurs also presumably in a basis of
co\-he\-rent states \cite{11r} although the validity of the present analysis
remains uncertain in that case, as well as its
relation with classical behavior \cite{12r} . \par

Finally, the increase of entropy resulting from irreversibility is mainly
found in the increase of $- Tr (\rho_c \ {\rm Log} \ \rho_c)$ when $\rho_c$
becomes approximately diagonal. \\

\noindent {\large\bf Acknowledgements} \par
I benefitted from useful suggestions by Roger Balian in the early stages of
this research.

\newpage
\def\labelenumi{[\arabic{enumi}]}
\noindent
{\large\bf Footnotes and References}
\begin{description}
\item[(a)]
Eq. (\ref{12e}) was first obtained as a consequence of a guess with
{\it a posteriori} justification \cite{7r}. In the same paper, I also
attempted its derivation by the projection method, although
making unfortunately a few self-compensating errors. No
applications except trivial ones were given.
\end{description}

\ben
  \item\label{1r} M. Brune, E. Hagley, J. Dreyer, X. Ma{\^\i}tre, C.
Wunderlich, J. M. Raimond, S. Haroche, Phys. Rev. Lett. {\bf 77}, 4887
(1996).
 
\item\label{2r} S. Nakajima, Progr. Theor. Phys. {\bf 20}, 948 (1958).
 
\item\label{3r} R. Zwanzig, Lect. Theor. Phys. (Boulder) {\bf 3}, 106
(1960)~; Physica {\bf 30}, 1109 (1964).

\item\label{4r} F. Haake, Springer Tracts in Modern Physics {\bf 66}, 98
(1973).

\item\label{5r}R. Balian, Y. Alhassid, H.
Reinhardt, Phys. Reports {\bf 131}, 1 (1986).

   \item\label{6r} D. Giulini, E. Joos, C. Kiefer, J. Kupsch, O.
Stamatescu, H. D. Zeh, {\it De\-co\-he\-ren\-ce and the Appearance of a
Classical World in Quantum Theory}, Springer, Berlin, 1996, Chapter 7 by J.
Kupsch.

\item\label{7r} R. Omn\`es, Phys. Rev. {\bf A56}, 3383 (1997).

   \item\label{8r} W. H. Zurek, Phys. Rev. {\bf D26}, 1862 (1982).

\item\label{9r} A. O. Caldeira, A. J. Leggett, Physica {\bf A121}, 587
(1983).
   \item\label{10r} E. Joos, H. D. Zeh, Z. Phys. {\bf B59}, 239 (1985).

\item\label{11r} W. H. Zurek, Prog. Theor. Phys. {\bf 61}, 281 (1993).

\item\label{12r} M. Gell-Mann, J. B. Hartle, Phys. Rev. {\bf D47}, 3345 (1993).
\een

\end{document}